\documentclass{iau}
\usepackage{graphicx,natbib}
 
\newcommand{\apj}{ApJ}           % Astrophysical Journal
\newcommand{\mnras}{MNRAS}       % Monthly Notices of the RAS

\newcommand{\aap}{A\&A}
\newcommand{\araa}{ARA\&A}

\newcommand{\pasp}{PASP}
\title{OMEGA: OSIRIS Mapping of Emission-Line Galaxies in A901/2}

\author[Bruno~Rodr\'iguez~del~Pino~et~al.]{Bruno~Rodr\'iguez~del~Pino$^{1}$, Ana~L.~Chies-Santos$^{1}$, Alfonso~Arag\'on-Salamanca$^{1}$, Steven~P.~Bamford$^{1}$, Meghan~E.~Gray$^{1}$}

\affiliation{$^1$School of Physics and Astronomy, The University of Nottingham, \\
University Park, Nottingham, NG7 2RD, UK \\
email: {\tt ppxbr@nottingham.ac.uk} \\}

\pubyear{2014}
\volume{309}
\jname{Galaxies in 3D across the Universe}
\editors{B. L. Ziegler, F. Combes, H. Dannerbauer, M. Verdugo, eds.}

\begin{document}

\maketitle

\begin{abstract}
 This work presents the first results from an ESO Large Programme carried out using the OSIRIS instrument on the 10m GTC telescope (La Palma). We have observed a large sample of galaxies in the region of the Abell 901/902 system (z$\sim$0.165) which has been extensively studied as part of the STAGES project. 
We have obtained spectrally and spatially resolved H-alpha and [NII] emission maps for a very 
large sample of galaxies covering a broad range of environments. The new data are combined with 
extensive multi-wavelength observations which include HST, COMBO-17, Spitzer, Galex and XMM imaging to study star formation and AGN activity as a function of environment and galaxy properties such as luminosity, mass and morphology. The ultimate goal is to understand, in detail, the effect of the environment on star formation and AGN activity. 
\end{abstract}

\firstsection
\section{Introduction}

The properties of the galaxies change as a function of environment. Galaxies in low-density regions tend to be blue, star-forming objects with disk morphology, whereas those living in denser regions tend to be red, passive objects with more spheroidal morphologies (\citealt{dressler80, bamford09}). The environment also seems to affect the probability of a galaxy hosting an Active Galactic Nucleus (AGN), at least in the high-luminosity end (\citealt{k04, popesso06}). By looking at a multi-cluster system at z~$\sim$~0.165, our main goal is to understand better the role played by the environment in the transformation of galaxies from actively star-forming to passive, with the subsequent morphological change, and the relation between AGN activity and environment. 

\section{OMEGA survey}

The OSIRIS Mapping of Emission-Line Galaxies in A901/2 (OMEGA) survey has been designed to study AGN and star formation activity as a function of environment in the A901/2 multi-cluster system at z$\sim$0.165. This system, which spans$\sim$0.5 $\times$ 0.5 deg$^2$ ($\sim$5x5~Mpc$^2$) contains a wide variety of environments and has been extensively studied by the STAGES collaboration (\citealt{gray09}). Besides that, there is already data from this system in different wavelengths, including \textit{XMM-Newton}, \textit{GALEX}, \textit{Spitzer}, 2dF, GMRT, Magellan, and the 17-band COMBO-17 photometric redshift survey. 

Despite the wealth of data aleready available, it is crucial to have optical diagnostics such as emission line spectra to really understand the roles of obscured and unobscured star formation, and the fraction of low-luminosity AGN with no X-Ray emission.
To obtain these missing optical diagnostics we have used the OSIRIS instrument in Narrow-band Tunable Filter mode at the GTC to map the H$\alpha$ and [N II] emission lines in the A901/2 multi-cluster system. These two lines contain a lot of information, since the H$\alpha$ line is a very good star formation indicator (\citealt{ken98}), and in combination with the [N II] line can provide diagnostics about the AGN activity in the galaxies, using the WHAN diagram introduced by \citealt{cid10}. 

To map these two emission lines in the whole system, we divided the $\sim$0.5 $\times$ 0.5 deg$^2$ field into 20 circular fields of 8 arcmin diameter each (OSIRIS field of view). For each field a series of `monochromatic' images was taken with a fixed wavelength step between them. For an optimal deblending of the two lines this step was set to 7\AA. The wavelength across the field of view also changes as a function of distance to the optical center, following expression (29) in \citealt{gonzalez14}. In this way, H$\alpha$ and [N II] are mapped by taking sufficient steps in wavelength based on the mean redshift of each field. In this work we present the results from the analysis of the two densest fields.
\begin{figure}[h]
\begin{center}
\includegraphics[width=0.7\columnwidth]{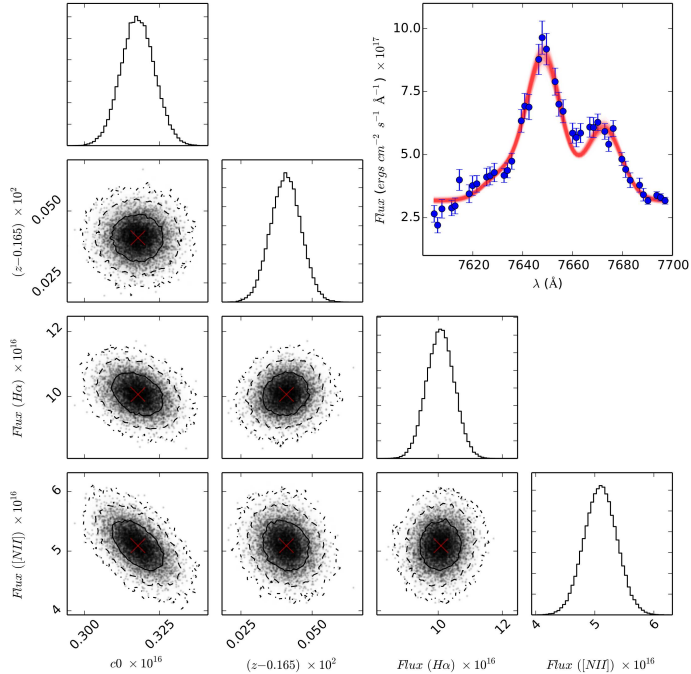}
\caption{Parameter-parameter likelihood distributions with one- (solid), two- (dashed) and three- (dash-dotted) sigma contours obtained for the four-parameter model of the spectrum of galaxy 42713. On the top right we show the spectrum together with the model using the median values of the parameters' probability distribution, which are also plotted as a red cross on the two-dimensional distributions.}
\label{mcmc}
\end{center}
\end{figure}
\section{Analysis and first results}
To build up the low-resolution spectra out of these images we performed aperture photometry using two different apertures: one matching the seeing of the images and another one encompassing the full extent of the galaxies. An example of one of these spectra is shown in Figure \ref{mcmc}.
As our main goal is to measure the fluxes in the two lines we need to model our data. Our model is a composite of three gaussians (because there is another weaker [N II] line at $\lambda$) and the continuum.

To exploit all the information contained in our data we use an algorigthm based on Markov Chain Monte Carlo (MCMC) techniques, which not only provides the maximum likelihood of the model parameters but also generates their full probability densitiy distributions. In our case we employ the software {\tt emcee} (\citealt{mcmcham}). An example of the performance of the algorithm is shown in figure \ref{mcmc}. From the output of the MCMC runs we selected a sample based on the probability of a detection in H$\alpha$, and a redshift consistent with one or the two lines being within the wavelength range.

\begin{figure*}
\begin{center}
\includegraphics[width=0.7\columnwidth]{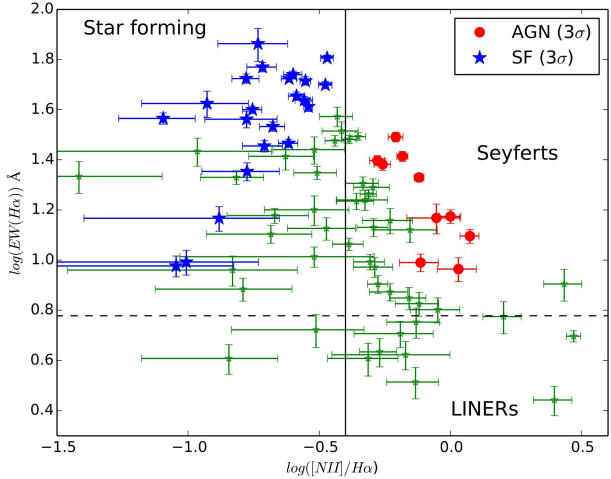}
\caption{A diagnostic plot of [N II]/H$\alpha$ vs.\@ EW (H$\alpha$), the so-called WHAN diagram (\citealt{cid10}). We only show detections, using $R_{PSF}$ aperture measurements, where both H$\alpha$ and [N II] fall in the OMEGA probed wavelength range: a total of 57 ELGs. From these we find 9 3$\sigma$-detected AGN hosts and 17 SF galaxies.}
\label{whan}
\end{center}
\end{figure*}

\begin{figure*}
\begin{center}
\includegraphics[width=0.7\columnwidth]{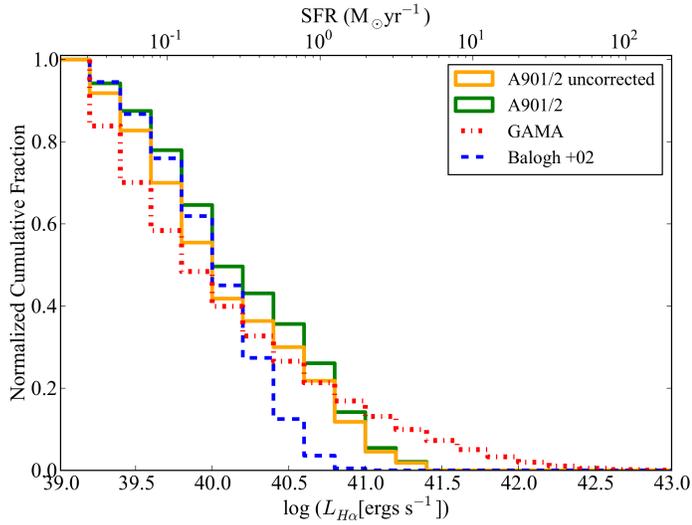}
\caption{The cumulative H$\alpha$ LF for the highest density regions of A901/2 (both corrected and uncorrected for contamination and completeness). For comparison we show the H$\alpha$ LF from the cluster A1689 at $z=0.18$ (\citealt{balogh02}) and the field from the GAMA survey at $z=0.1$--$0.2$ (\citealt{guna13}), according to the legend. The studied regions of OMEGA have SFRs that fall in between those of the cluster A1689 and the field.}
\label{half}
\end{center}
\end{figure*}

From the analysis of the two densest fields we use the WHAN diagnostic diagram (\citealt{cid10}) in Figure \ref{whan} to evaluate the probability of the galaxies hosting an AGN, star formation or both. By looking at the spatial distribution of these sources we find that both AGN and star-forming galaxies tend to avoid the densest regions. The H$\alpha$ Luminosity Function, shown in Figure \ref{half}, when compared with other studies, shows fewer high star-forming objects than the field, but more than in dense clusters. Using the SFR vs Stellar Mass relation, there is also evidence of a widespread suppression of the star formation in the cluster galaxies compared with the field ones. Besides that, our SFR estimates based on the H$\alpha$ flux agree well with other SFR indicators. 

\section{Conclusions and future work}

The analysis of only one tenth (2 fields out of 20) of the data of the OMEGA survey has shown that Tunable Filter observations are a very powerful tool to obtain emission line fluxes in a large number of galaxies in single redshift windows. The results found in this work make the analysis of the whole data set very promising as we will be able to show the environmental effects on AGN and star formation activity in $\sim$ 1000 objects in the multi-cluster system at z=0.165. 

A complete version of the data reduction, analysis and first results of this survey can be found in Chies-Santos et al. in prep.

\end{document}